\documentclass[conference]{IEEEtran}
\IEEEoverridecommandlockouts

\usepackage{cite}
\usepackage{amsmath,amssymb,amsfonts}
\usepackage{algorithmic}
\usepackage{graphicx}
\usepackage{textcomp}
\usepackage{xcolor}

\usepackage{booktabs}    
\usepackage{url}     
\usepackage{tikz}        
\usetikzlibrary{shapes,arrows,positioning,calc,fit,arrows.meta}
\usepackage{placeins}
\def\BibTeX{{\rm B\kern-.05em{\sc i\kern-.025em b}\kern-.08em
    T\kern-.1667em\lower.7ex\hbox{E}\kern-.125emX}}

\newcommand{\cmark}{\ensuremath{\checkmark}}
\newcommand{\xmark}{\ensuremath{\times}}

\begin{document}

\title{OAK: Restart- and Age-Aware Scheduling for Distributed Machine Learning on Shared GPU Clusters}

\author{
\IEEEauthorblockN{Khaled Aljbab}
\IEEEauthorblockA{\textit{Department of Computer Science and Engineering} \\
\textit{Oakland University}\\
Rochester, MI, USA \\
aljbab@oakland.edu}
\and
\IEEEauthorblockN{Amine Barrak}
\IEEEauthorblockA{\textit{Department of Computer Science and Engineering} \\
\textit{Oakland University}\\
Rochester, MI, USA \\
aminebarrak@oakland.edu}
}

\maketitle

\begin{abstract}
Distributed machine learning increasingly runs on shared multi-tenant GPU clusters where contention and failures are routine. Goodput-driven schedulers maximise instantaneous throughput but treat accumulated waiting time and restart cost as second-class signals: long-waiting jobs are repeatedly deferred, and the minutes of checkpoint loading paid after each interruption are not folded back into allocation decisions.

We present OAK, a per-round mixed-integer linear scheduler whose composite utility adds two first-class terms to goodput: an age key that elevates jobs with high cumulative waiting time, and a decomposed restart factor that tracks productive training time and checkpoint overhead as separately measured quantities rather than a single aggregate ratio.

We evaluate OAK against four representative goodput- and fairness-driven baselines in a trace-driven simulator on a 12-GPU cluster across four data-parallel workloads with controlled Poisson failure injection, and reproduce the failure-mode improvement on a 4-V100 real-hardware cluster.

Under failure free conditions OAK matches the strongest goodput driven baseline within $4\%$. Under failures at rate $\lambda = 0.1$ it reduces mean job completion time (JCT) by $33.9$--$58.3\%$ and worst case (maximum) JCT by $58$--$76\%$ over its goodput driven foundation; the decomposed restart factor alone accounts for a $40$--$57\%$ mean and $52$--$76\%$ worst case JCT improvement against an aggregate restart estimator used in prior goodput driven schedulers. Per round decision latency is $7.65$\,ms, three orders of magnitude faster than evolutionary search alternatives at default settings.

\end{abstract}


\section{Introduction}
\label{sec:intro}

Deep-learning training has become a dominant workload on shared multi-tenant GPU clusters~\cite{weng2022mlaas}, and its operational cost has grown with model scale. A modern training run can occupy dozens of GPUs for days or weeks, so the cluster scheduler's per-round allocation decisions directly determine job completion time (JCT)---how quickly users get results---how predictable tail latency looks for jobs bound by service-level agreements (SLAs), and how much of the cluster's purchased capacity becomes useful work. Scheduling distributed machine-learning (DML) workloads differs from scheduling traditional batch jobs in two ways that make it substantially harder. First, individual jobs are long-lived and checkpoint-coupled: preempting a job is not free; it pays a real cost in time-to-resume from the last checkpoint, and that cost grows with model size~\cite{maurya2024datastates,gandhi2024moetion}. Second, jobs are failure-prone: at the scale of contemporary DML training, hardware and software faults are routine rather than exceptional~\cite{he2023unicron}. A scheduler that treats every running job as immortal will repeatedly absorb restart cost that better allocations could have avoided.

Production measurements quantify both pressures. The MLaaS-in-the-Wild study of large-scale GPU clusters~\cite{weng2022mlaas} reports that contention and heterogeneity dominate real-world workloads, with a non-trivial fraction of GPU-time consumed by failed or restarted jobs. Failure-recovery work for large models~\cite{he2023unicron,maurya2024datastates,gandhi2024moetion} similarly documents that recovery overhead scales poorly with model size and that frequent interruption of long-running training is a first-class operational concern rather than an edge case.

Existing schedulers address parts of this problem but not the combination. Goodput-driven schedulers such as Sia~\cite{jayaram2023sia} and Pollux~\cite{qiao2021pollux} re-decide GPU allocations every round to maximise throughput, but they do not surface accumulated waiting time as a first-class signal, and Pollux's restart-aware factor folds productive training time and checkpoint overhead into a single aggregate estimator rather than tracking them separately. Fairness-driven schedulers such as Gavel~\cite{narayanan2020heterogeneity} and DRF~\cite{ghodsi2011dominant} target finish-time fairness or dominant-resource fairness but do not internalise restart cost in their allocation objectives. Straggler-aware execution~\cite{ozfatura2020straggler,amiri2019computation,adikari2024exploiting} addresses worker-level latency variability rather than the cluster-allocation question. None of these lines combines goodput, accumulated waiting, and measured restart cost in a single per-round objective.

We propose OAK, a composite-utility scheduler that addresses both gaps within Sia's goodput-driven foundation. OAK adds two first-class signals to the per-round score: an age key that encodes accumulated queue waiting as a continuous bounded multiplier, and a decomposed restart factor that tracks productive training time and checkpoint overhead as separately measured quantities rather than collapsing them into a single ratio. The combined score is plugged into the same per-round mixed-integer linear program Sia already uses, keeping the optimiser linear.

This paper makes the following contributions:

\begin{itemize}
    \item We integrate goodput, restart cost, and accumulated waiting time in a single per-round mixed-integer linear program (MILP) with a sub-10\,ms decision time at the 12-GPU scale. The same composite utility delivers Sia-beating failure resilience at this latency, a combination none of the four baselines achieves.
    \item We track productive training time and checkpoint/restore overhead as separately measured quantities, in contrast to Pollux's aggregate multiplicative estimator. The decomposition alone yields a $40$--$57\%$ mean JCT reduction and a $52$--$76\%$ worst case (maximum) JCT reduction at $\lambda = 0.1$.
    \item We report a trace-driven simulator comparison (OAK, Sia, Pollux, Gavel, DRF) on a 12-GPU cluster across four data-parallel workloads, supplemented by a 4-V100 real-hardware validation.\footnote{Code, configurations, and traces for reproducing the results are available at \url{https://sites.google.com/oakland.edu/oak-icpads2026/home}.}
\end{itemize}

The remainder of the paper is organised as follows. Section~\ref{sec:related} reviews prior scheduling and failure recovery work. Section~\ref{sec:oak_design} presents OAK's composite utility and per round MILP. Section~\ref{meth} details the simulator, workloads, and baselines. Section~\ref{sec:results} reports the empirical results, including the OAK versus Sia and Pollux comparisons, the decomposed restart factor ablation, and a 4-V100 real hardware validation. Section~\ref{sec:threats} discusses threats to validity, and Section~\ref{sec:conclusion} concludes.

\section{Related Work}
\label{sec:related}

\subsection{Cluster Schedulers for Deep-Learning Training}
\label{sec:rel_schedulers}

Cluster-level scheduling for deep-learning workloads has been studied along two axes: allocation efficiency and goodput- or fairness-driven optimisation. Allocation-efficient schedulers such as Optimus~\cite{peng2018optimus}, Aryl~\cite{li2022aryl}, DL2~\cite{peng2021dl2}, Titan~\cite{gao2022titan}, and Lucid~\cite{hu2023lucid} target specific operational concerns (predictive scaling, elasticity, GPU utilisation) but do not jointly optimise goodput, waiting time, and restart cost. The goodput- and fairness-driven line of work is the closest to OAK and includes DRF~\cite{ghodsi2011dominant}, Gavel~\cite{narayanan2020heterogeneity}, Pollux~\cite{qiao2021pollux}, and Sia~\cite{jayaram2023sia}. Production GPU-cluster measurements~\cite{weng2022mlaas} motivate this line by showing that contention and heterogeneity dominate real-world workloads, so a single scheduler must reconcile efficiency and fairness rather than optimise either alone.

Sia~\cite{jayaram2023sia} provides our architectural foundation: configuration-based, heterogeneity-aware allocation that re-decides each job's GPU assignment every round to maximise goodput, formulated as a per-round mixed-integer linear program. Pollux~\cite{qiao2021pollux} shares the goodput objective but solves the joint allocation through NSGA II evolutionary search and includes a multiplicative restart-aware factor $\max(A_i - n_i \cdot P_{\mathrm{restart}}, 0)/(A_i + P_{\mathrm{restart}})$ that aggregates progress and overhead into a single estimator. Gavel~\cite{narayanan2020heterogeneity} treats fairness through finish-time-fairness or max-sum-throughput policies but does not internalise restart cost or accumulated waiting time. DRF~\cite{ghodsi2011dominant} allocates dominant-resource shares without a goodput model.

OAK extends this line by adding waiting time and restart cost as first-class signals inside a single per-round score $(G \cdot r)^p \cdot K$: the age key $K_i$ encodes waiting time as a bounded multiplier, and the restart factor $r_i$ replaces Pollux's aggregate estimator with directly measured progress and checkpoint time. Table~\ref{tab:scheduler-novelty} summarises these distinctions; none of the four baselines combines goodput with both signals in one objective.

\begin{table}[t]
  \centering
  \caption{Per-round objectives across baselines and OAK. \cmark/\xmark\ mark whether waiting time and restart cost are first-class signals; ``Aggregate'' means progress and overhead are folded into one ratio.}
  \label{tab:scheduler-novelty}
  \scriptsize
  \setlength{\tabcolsep}{3pt}
  \renewcommand{\arraystretch}{1.15}
  \begin{tabular}{@{}l p{1.85cm} p{0.95cm} p{1.4cm} l@{}}
    \toprule
    \textbf{Scheduler} &
      \shortstack[l]{\textbf{Optimisation}\\\textbf{objective}} &
      \shortstack[l]{\textbf{Waiting}\\\textbf{time}} &
      \shortstack[l]{\textbf{Restart}\\\textbf{cost}} &
      \textbf{Solver} \\
    \midrule
    DRF~\cite{ghodsi2011dominant}
      & Dominant-resource fairness
      & \xmark
      & \xmark
      & Max-min \\
    Gavel~\cite{narayanan2020heterogeneity}
      & FTF / max-sum throughput
      & Indirect (FTF only)
      & \xmark
      & LP \\
    Sia~\cite{jayaram2023sia}
      & Goodput
      & \xmark
      & Aggregate
      & ILP \\
    Pollux~\cite{qiao2021pollux}
      & Goodput + statistical efficiency
      & \xmark
      & Aggregate
      & NSGA II \\
    \midrule
    \textbf{OAK} (this work)
      & Composite $(G\!\cdot\! r)^p\!\cdot\! K$
      & \cmark\ ($K_i$)
      & \cmark\ ($r_i$, decomposed)
      & ILP \\
    \bottomrule
  \end{tabular}
\end{table}

\subsection{Restart Cost and Recovery in DML Training}
\label{sec:rel_recovery}

A complementary line of systems work targets the cost of an individual restart by improving checkpoint and recovery mechanisms at the runtime layer. MoEtion~\cite{gandhi2024moetion} reduces checkpoint overhead for mixture-of-experts training, while MoC-System~\cite{cai2025moc} and the asynchronous checkpointing approach for large language models~\cite{maurya2024datastates} drive down the wall-clock cost of saving and restoring model state during training. At the failure-recovery layer, Unicron~\cite{he2023unicron} targets large-model training by overlapping recovery with surviving-replica progress, and adaptive fault-tolerance schemes for cloud LLM training~\cite{jin2025adaptive} combine partial reload with selective re-execution to amortise recovery cost. 

A second related thread operates below the scheduler at the worker and network level. Straggler-aware execution~\cite{ozfatura2020straggler,amiri2019computation,adikari2024exploiting} addresses latency variability among workers during a single training step rather than across allocation decisions. Network and transport designs for distributed training~\cite{xia2019rethinking,viswanathan2020network,cardoso2025no} optimise gradient synchronisation and in-network aggregation paths but leave the per-round allocation policy untouched.

These approaches are largely orthogonal to cluster scheduling: they reduce the cost paid for a single recovery event but do not feed restart-induced inefficiency back into the allocation objective. A scheduler that ignores accumulated checkpoint time will keep preempting the same job, paying the recovery cost repeatedly.

OAK is complementary to runtime-level checkpoint and recovery systems. Rather than reducing the cost of a single restart, the decomposed restart factor $r_i$ feeds accumulated checkpoint time back into the per-round score, steering the scheduler away from configurations that would generate further restart cost in the first place. 

\section{Design of OAK}
\label{sec:oak_design}

OAK is a scheduler for distributed machine learning (DML) clusters that adds two extensions to Sia's existing design~\cite{jayaram2023sia}: an \emph{age key} that prioritises jobs with long accumulated waiting time, and a \emph{restart factor} that penalises configurations whose effective progress is eroded by checkpoint and recovery overhead. Both terms are folded into the same per-round scoring function alongside goodput, so the scheduler weighs efficiency, waiting time, and restart cost together in one decision.

\paragraph{Relationship to Sia.}
OAK extends Sia rather than replacing it. The only thing OAK changes is the per-pair (job, configuration) score that the ILP maximises: it adds an age-key term $K_i$ and a restart-factor term $r_i$ to the goodput term $G_{ij}$ that Sia already uses. Every other Sia component is reused without modification:
\begin{itemize}
    \item The per-job set of candidate configurations $C_i$ (replica counts and GPU types Sia considers feasible for job $i$);
    \item The trace-based goodput estimate $G_{ij}$ that predicts each (job, configuration) throughput;
    \item The placement step that, after the optimiser decides how many GPUs each job receives, maps that count to specific GPU IDs on specific nodes;
    \item The optimisation back-end, the same mixed-integer linear program is built with the CVXPY modelling library~\cite{diamond2016cvxpy} and solved with the GLPK\_MI integer solver, both of which are open source and standard in the operations-research toolchain.
\end{itemize}
Because OAK and Sia differ only in the score, any JCT difference between them in our experiments can be attributed to the age key and the restart factor.

\subsection{Scheduling Model and Notation}
\label{sec:oak_model}

OAK schedules in \emph{rounds}. Round $t$ begins when a fixed timer fires (every 60\,s in our experiments) or when the active job set changes (a job arrives, completes, or is preempted). At the start of each round the scheduler reads the current cluster state, picks an allocation, and applies it until the next round.

We use the following notation throughout.

\begin{itemize}
    \item \textbf{Cluster.} $G = \{g_1, \dots, g_m\}$ is the set of GPU types present in the cluster, and $R_g$ is the number of GPUs of type $g$. Each GPU is assigned to at most one job per round (no time-sharing within a round).
    \item \textbf{Active jobs.} $J_t$ is the set of jobs active at round $t$ --- those that have been submitted but have not yet completed.
    \item \textbf{Configurations.} For each job $i$, the candidate set $C_i$ lists the configurations the job can run. A configuration $j \in C_i$ specifies a target GPU type, a replica count, and the per-type GPU demand $d_g(i, j)$.
\end{itemize}

Before the round-level optimisation runs, OAK filters $C_i$ to drop configurations that exceed per-GPU memory budgets or violate workload-specific constraints. As a result, the per-pair score $U_{ij}$ and the ILP structure depend only on the surviving configurations; they do not depend on which jobs happen to be active in a given round.

Two per-job temporal counters drive OAK's new utility terms:
\begin{itemize}
    \item $T_{\mathrm{queue},i}(t)$: how long job $i$ has been waiting without GPUs. Used by the age key to boost long-waiting jobs.
    \item $T_{\mathrm{ckpt},i}(t)$: how long job $i$ has spent in checkpoint or restore. Used by the restart factor to penalise jobs whose progress is eroded by overhead.
\end{itemize}
The job's age is $A_i(t)$, and its productive training time is $T_{\mathrm{progress},i}(t) = A_i(t) - T_{\mathrm{ckpt},i}(t)$.

\subsection{Composite Utility}
\label{sec:oak_utility}

For each pair (job $i$, feasible configuration $j$), OAK assigns a score
\begin{equation}
U_{ij}(t) = \big(G_{ij} \cdot r_i(t)\big)^{\,p} \cdot K_i(t),
\label{eq:oak-utility}
\end{equation}
where $G_{ij}$ is the goodput estimate inherited from Sia, $r_i(t) \in (0,1]$ is the restart factor (down-weights jobs eroded by checkpoint and recovery overhead), and $K_i(t) \ge 1$ is the age key (boosts long-waiting jobs).

The fairness exponent $p$ shapes how much additional goodput contributes to the score: at $p = 0.5$ (used throughout) doubling goodput multiplies the score by only $\sqrt{2}$, so the scheduler prefers high-goodput allocations without chasing them aggressively. This produces \emph{soft} fairness rather than strict max-min behaviour. Negative values of $p$ cause the score $(G_{ij} \cdot r_i)^p$ to diverge when $r_i \to 0$ (a job dominated by checkpoint time), so we restrict $p \ge 0$ here and leave strictly fair regimes for future work. Assigning separate exponents to $G_{ij}$ and $r_i$ is a natural generalisation; we couple them here because both express per-round execution efficiency, and leave independent shaping to future work.

The age key $K_i$ sits \emph{outside} the power: this keeps the waiting-time boost as a bounded multiplicative signal whose magnitude is independent of the goodput-shaping exponent $p$. Placing $K_i$ inside the power would couple its effect to $p$ and require re-tuning the age-key parameters every time $p$ changes.

\paragraph{Per-round optimisation.}
At each round, OAK chooses binary variables $x_{ij}(t) \in \{0,1\}$, where $x_{ij} = 1$ means job $i$ is assigned configuration $j$:
{\small
\begin{align}
\max_{\{x_{ij}\}} \quad
& \sum_{i \in J_t}\sum_{j \in C_i} x_{ij}\, U_{ij}(t) + \mu_{\mathrm{na}} \sum_{i \in J_t} \Big(1 - \sum_{j \in C_i} x_{ij}\Big) \label{eq:oak-opt} \\
\qquad\text{s.t.}\qquad
& \quad \sum_{j \in C_i} x_{ij} \le 1, && \forall i \in J_t \nonumber \\
& \quad \sum_{i,j} x_{ij}\, d_g(i,j) \le R_g, && \forall g \in G \nonumber
\end{align}
}
The constant $\mu_{\mathrm{na}} = 1.1$ acts as a utility threshold for allocation. The solver commits GPUs to job $i$ only if at least one of its feasible utilities $U_{ij}$ exceeds $\mu_{\mathrm{na}}$; if every $U_{ij}$ for that job falls below the threshold, leaving the job idle is more attractive to the objective than any of its candidate allocations, because the unallocated branch contributes $\mu_{\mathrm{na}}$ per idle job. The power transformation $(G_{ij} \cdot r_i)^p$ from Eq.~\ref{eq:oak-utility} is precomputed before the optimisation step, so the solver receives a purely linear integer program over $|J_t| \cdot \max_i |C_i|$ binary variables.


\subsection{Restart Factor and Age Key}
\label{sec:oak_terms}

\paragraph{Restart factor.}
OAK's \emph{decomposed} restart factor is
\begin{equation}
r_i(t) = \frac{\max\!\big(T_{\mathrm{progress},i}(t),\, 0\big)}{T_{\mathrm{progress},i}(t) + T_{\mathrm{ckpt},i}(t) + P_{\mathrm{restart},i}},
\label{eq:oak-restart}
\end{equation}
where $T_{\mathrm{progress},i}(t) = A_i(t) - T_{\mathrm{ckpt},i}(t)$ is the job's productive training time (Section~\ref{sec:oak_model}) and $P_{\mathrm{restart},i}$ is the per-application restart penalty (covering checkpoint loading, state reconstruction, and warm-up). Checkpoint and recovery time freezes the numerator (the job ages without training) while inflating the denominator through $T_{\mathrm{ckpt},i}$; both effects lower $r_i$. For freshly submitted jobs, where $A_i(t) < P_{\mathrm{restart},i}$, the formula is overridden to $r_i = 1$ so that the score is not penalised before the job has had a chance to make progress. Note that $r_i$ penalises restart-heavy \emph{configurations}, not users: a job repeatedly hit by infrastructure faults loses score through $r_i$, but its queue waiting simultaneously accumulates in $K_i$, which partially offsets the penalty. Distinguishing user-caused from system-caused failures in the score is left to future work.

\paragraph{Contrast with Pollux.}
Pollux~\cite{qiao2021pollux} down-weights restart-heavy jobs through an aggregate factor that \emph{estimates} the time lost to restarts as $n_i \cdot P_{\mathrm{restart}}$ (restart count times a constant penalty). The difference is precision: this approximation accumulates estimation error as restarts pile up, whereas OAK's directly measured $T_{\mathrm{ckpt},i}$ remains exact at any failure rate. Section~V-D shows the two formulas side by side and isolates the empirical effect of this design choice.

\begin{figure}[t]
\centering
\resizebox{0.96\columnwidth}{!}{%
\begin{tikzpicture}[
  font=\footnotesize,
  flowstep/.style={
    draw, rounded corners=2pt, align=left, inner sep=4pt,
    minimum width=7.4cm, minimum height=0.7cm
  },
  oak/.style={
    draw, rounded corners=2pt, align=left, inner sep=4pt,
    minimum width=7.4cm, minimum height=0.7cm,
    fill=blue!8, very thick
  },
  start/.style={
    draw, rounded corners=8pt, align=center, inner sep=4pt,
    minimum width=5.4cm, minimum height=0.6cm,
    fill=gray!10
  },
  arr/.style={thick, -{Latex[length=2.5mm]}}
]
\node[start] (start) {\textbf{Round} $t$: scheduling event (60\,s timer or job arrival/completion)};
\node[flowstep, below=0.5cm of start] (s1)
  {\textbf{1. State update} (Queue Manager): refresh $T_{\mathrm{queue},i}$, $T_{\mathrm{ckpt},i}$ from simulator state};
\node[flowstep, below=0.45cm of s1] (s2)
  {\textbf{2. Goodput estimate} (Profiler $\to$ Estimator): per (job\,$i$, config\,$j$) compute $G_{ij}$ from traces};
\node[oak, below=0.45cm of s2] (s3)
  {\textbf{3. Restart factor [OAK]}: $r_i = \dfrac{T_{\mathrm{progress},i}}{T_{\mathrm{progress},i}+T_{\mathrm{ckpt},i}+P_{\mathrm{restart},i}}$};
\node[oak, below=0.45cm of s3] (s4)
  {\textbf{4. Age key [OAK]}: $K_i = \min\!\big(\exp(\alpha\,T_{\mathrm{queue},i}),\;K_{\max}\big)$};
\node[oak, below=0.45cm of s4] (s5)
  {\textbf{5. Composite utility [OAK]}: $U_{ij}(t) = \big(G_{ij}\!\cdot r_i(t)\big)^{p}\cdot K_i(t)$};
\node[flowstep, below=0.45cm of s5] (s6)
  {\textbf{6. ILP solve} (GLPK\_MI): $\max\sum_{i,j}x_{ij}\,U_{ij}$ s.t. capacity \& architecture feasibility};
\node[flowstep, below=0.45cm of s6] (s7)
  {\textbf{7. Map to GPUs} (Placer): exclusivity check, repair, logical-to-physical GPU mapping};
\node[flowstep, below=0.45cm of s7] (s8)
  {\textbf{8. Apply allocation} (Adaptive Executors): preempt / start / resize jobs, trigger checkpoint+resume cycle};
\node[start, below=0.5cm of s8] (next) {\textbf{Wait} for next round / event};

\draw[arr] (start) -- (s1);
\draw[arr] (s1) -- (s2);
\draw[arr] (s2) -- (s3);
\draw[arr] (s3) -- (s4);
\draw[arr] (s4) -- (s5);
\draw[arr] (s5) -- (s6);
\draw[arr] (s6) -- (s7);
\draw[arr] (s7) -- (s8);
\draw[arr] (s8) -- (next);

\end{tikzpicture}%
}
\caption{OAK's per-round scheduling workflow. Blue steps (3--5) are OAK's additions; the rest are inherited from Sia.}
\label{fig:oak-workflow}
\end{figure}
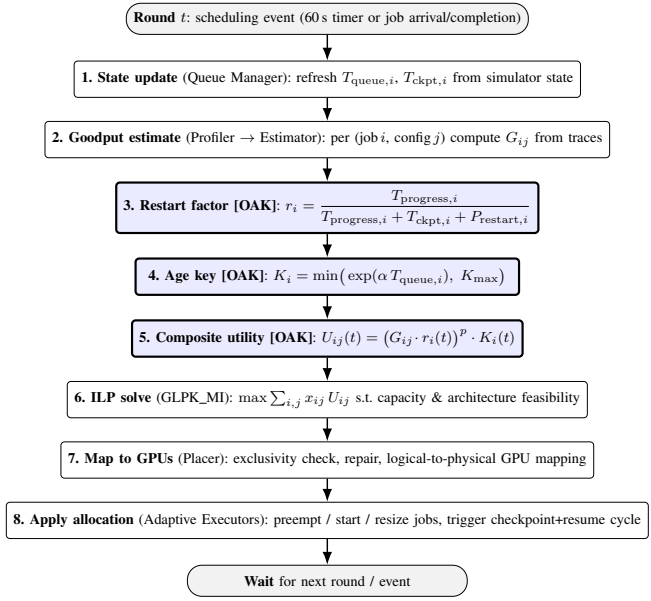

\paragraph{Age key.}
The age key turns the queue-waiting counter $T_{\mathrm{queue},i}(t)$ into a multiplicative score boost that grows with waiting time but never exceeds a fixed cap:
\begin{equation}
K_i(t) = \min\!\Big(\exp\big(\alpha\, T_{\mathrm{queue},i}(t)\big),\; K_{\max}\Big),
\label{eq:oak-age}
\end{equation}
The exponential term raises a job's score the longer it waits, while the upper bound $K_{\max}$ prevents a single starved job from dominating the cluster. We use $\alpha = 0.01$ and $K_{\max} = 100$ in all experiments. These defaults give waiting jobs a gradual priority boost without letting any single job dominate the cluster. The cap serves two purposes. First, it bounds the worst-case throughput cost of prioritising any single job: without a cap, a sufficiently aged job would dominate the round regardless of efficiency, and a user could exploit this by submitting jobs designed to accumulate waiting time. Second, $K_{\max} = 100$ is deliberately above the values reached in our workloads: it acts as a robustness guard rather than an active tuning knob, so operators need not recalibrate it per workload. We use an exponential rather than linear aging law so that priority accelerates with starvation: a linear key raises long- and briefly-waiting jobs at the same rate, whereas the exponential form keeps briefly delayed jobs near $K_i \approx 1$ while sharply promoting chronically starved ones before the cap binds.

\paragraph{Worked example.}
Consider three jobs contending for one 4 GPU configuration: $J_1$ (goodput $G_1 = 2.0$, $r_1 = 1.0$, $K_1 \approx 1.06$), $J_2$ (mid training ImageNet, $G_2 = 2.2$, $r_2 = 0.50$ after two failures, $K_2 \approx 1.06$), and $J_3$ (chronically starved, $G_3 = 1.4$, $r_3 = 1.0$, $K_3 = \exp(0.01\cdot 400) \approx 54.6$). Sia picks $J_2$ by goodput alone. OAK's composite utility (Eq.~\ref{eq:oak-utility}, $p = 0.5$) evaluates to $U_1 \approx 1.50$, $U_2 \approx 1.11$, $U_3 \approx 64.6$, so with $\mu_{\mathrm{na}} = 1.1$ the ILP admits $J_1$ and $J_3$ and de prioritises the restart heavy $J_2$, protecting a job whose budget is already burning while surfacing a chronically waiting one, neither of which a goodput only score can do.

\subsection{Per-Round Scheduling Workflow}
\label{sec:oak_workflow}

Figure~\ref{fig:oak-workflow} summarises the eight-step round. Steps 3--5 (blue) are OAK's additions; Steps 1--2 and 6--8 are inherited unchanged from Sia.

\paragraph{Solver-agnostic property.}
Steps 3--5 produce the score $U_{ij}$ and Steps 6--7 perform the optimisation; any optimiser that accepts a per-pair utility can replace Steps 6--7 without changing the score function, a deployment knob we revisit empirically in Section~V-E.

\section{Methodology}
\label{meth}

\subsection{Hardware and Simulator}
\label{sec:meth_hw}

All experiments run on a high performance computing cluster managed by SLURM. The cluster provides three GPU compute nodes (12 V100 GPUs in total); each node has 4 NVIDIA Tesla V100 16\,GB GPUs with NVLink, 192\,GB RAM, and 48 CPU cores at 2.10\,GHz, connected via HDR100 InfiniBand at 100\,Gbps.

The OAK simulator is built on the Sia simulator codebase~\cite{jayaram2023sia} and models job arrivals, allocation decisions, goodput estimation, checkpoint and restore dynamics, and queue management within a configurable cluster. The simulator supports pluggable scheduling policies, which enables A/B comparisons between OAK and baselines under identical workload and cluster conditions. Every scheduler we compare against (OAK, Sia, Pollux, Gavel, DRF) runs a new scheduling round once every 60 seconds.

Goodput estimates $G_{ij}$ that feed every scheduler are trace driven: per application step time and scaling traces collected on AWS T4 GPUs are inherited from the Sia codebase, and the simulator interpolates between trace points for unobserved (batch size, replica count) combinations. All scheduler comparisons in this paper are paired (same workload, same submission times, same Poisson failure stream, same trace based $G_{ij}$ values), so the relative JCT differences reported below are invariant to the simulator's absolute throughput calibration.

\subsection{Workloads}
\label{sec:meth_workloads}

Workloads are derived from real deep learning training profiles. They cover three job families: computer vision (CIFAR-10, ImageNet, YOLOv3), natural language processing (BERT), and speech (DeepSpeech2). Each job has a GPU demand (replica count), runtime dynamics taken from profiling traces, and per application checkpoint and restart behaviour modelled by the simulator. Job arrivals follow a time indexed submission process: each job has a submission timestamp and enters the queue at that time, with the trace timing chosen to induce overlap and contention so scheduling decisions have measurable impact.

We use four contention scenarios that escalate the queue pressure on the cluster: \texttt{workload-small} (4 jobs, light load, validates correctness), \texttt{workload-medium} (12 jobs, typical cluster usage), \texttt{workload-congestion} (24 jobs, resource scarcity), and \texttt{workload-stress} (40 jobs, the queue is consistently full).

\subsection{Composite Utility Solver}
\label{sec:meth_solver}

At each scheduling round OAK solves a mixed integer linear program (MILP). The composite utility uses a power transformation with $p = 0.5$ applied only to the goodput restart product; we precompute $\tilde{U}_{ij} = (G_{ij} \cdot r_i)^{p} \cdot K_i$ in NumPy before the values reach the solver, so the solver itself sees a purely linear objective over binary allocation variables $x_{ij} \in \{0,1\}$. Placing $K_i$ outside the power keeps the age boost independent of the fairness exponent $p$ and matches the utility definition in Section~\ref{sec:oak_design}. The MILP is formulated in CVXPY~\cite{diamond2016cvxpy} and solved with the GLPK\_MI backend, with an automatic fallback to CBC if GLPK\_MI errors. At our cluster scale (about 14 jobs, 10 configurations per job, 12 GPUs) per round solve times stay below 100\,ms, well within the 60 second scheduling interval.

The simulator filters infeasible configurations (those exceeding GPU memory budgets or violating workload-specific resource constraints) before they reach the solver. These checks operate on the input set $C_i$ (the feasible configurations for each job) rather than inside the optimisation, so the score and solver structure are unchanged.

\subsection{Failure Injection and Restart Factor}
\label{sec:meth_failures}

Failures are injected as a Poisson process at rate $\lambda$ per round per active job. When a job fails the simulator adds the per application restart penalty to a per job counter $T_{\mathrm{ckpt},i}$ that accumulates total time spent in checkpoint and restore. The simulator also tracks a queue waiting time $T_{\mathrm{queue},i}$ that grows when the job has no allocation \emph{and} when it is allocated but not making training progress (i.e., during a checkpoint or restore). Both counters feed OAK's restart factor and age key.

OAK's restart factor extends the aggregate reallocation form used by Sia and Pollux, $r_i = \max(A_i - n_i \cdot P_{\mathrm{restart}}, 0)/(A_i + P_{\mathrm{restart}})$, which \emph{estimates} the time lost across $n_i$ restarts as $n_i \cdot P_{\mathrm{restart}}$. OAK instead \emph{measures} the loss directly: the simulator-tracked counter $T_{\mathrm{ckpt},i}$ replaces the $n_i \cdot P_{\mathrm{restart}}$ estimate inside the decomposed factor of Section~\ref{sec:oak_terms}, so the penalty remains exact at any failure rate. For newly submitted jobs (where $A_i < P_{\mathrm{restart}}$) we set $r_i = 1$ so the score is not penalised before the job has had a chance to make progress.

The per-application restart penalties used throughout this paper are: CIFAR-10: $50$\,s; DeepSpeech2: $25$\,s; BERT: $120$\,s; YOLOv3: $80$\,s; ImageNet: $250$\,s; NCF: $25$\,s; GPT (pipeline-parallel): $30$\,s. A $30$\,s fallback is used for applications not in the map. These values are configured to reflect the relative recovery cost of each model family; Section~V-A quantifies the scheduler's sensitivity to this parameter.

\subsection{Baselines}
\label{sec:meth_baselines}

We compare OAK against four baseline schedulers, chosen to span the spectrum from pure fairness to pure efficiency:
\begin{itemize}
  \item \textbf{Sia}~\cite{jayaram2023sia}: the goodput optimised, heterogeneity aware ILP scheduler on which OAK is built (Section~\ref{sec:oak_design}). We run Sia with its default fairness exponent ($p = 0.5$).
  \item \textbf{Pollux}~\cite{qiao2021pollux}: NSGA II multi objective optimiser jointly considering goodput and statistical efficiency. We run Pollux at its published default NSGA II settings (population $100$, generations $100$) throughout.
  \item \textbf{Gavel}~\cite{narayanan2020heterogeneity}: a configurable, heterogeneity aware scheduler. We run Gavel with the \emph{finish-time fairness} (FTF) policy, in which allocations are chosen by a linear program that minimises the maximum ratio of each job's actual finish time to its proportional-share finish time. Gavel is deterministic.
  \item \textbf{DRF}~\cite{ghodsi2011dominant}: Dominant Resource Fairness; allocates GPU shares so that each job receives an equal fraction of its dominant resource. We use the standard formulation with no throughput optimisation; DRF is deterministic.
\end{itemize}

\begin{figure*}[t]
\centering
\includegraphics[width=\linewidth]{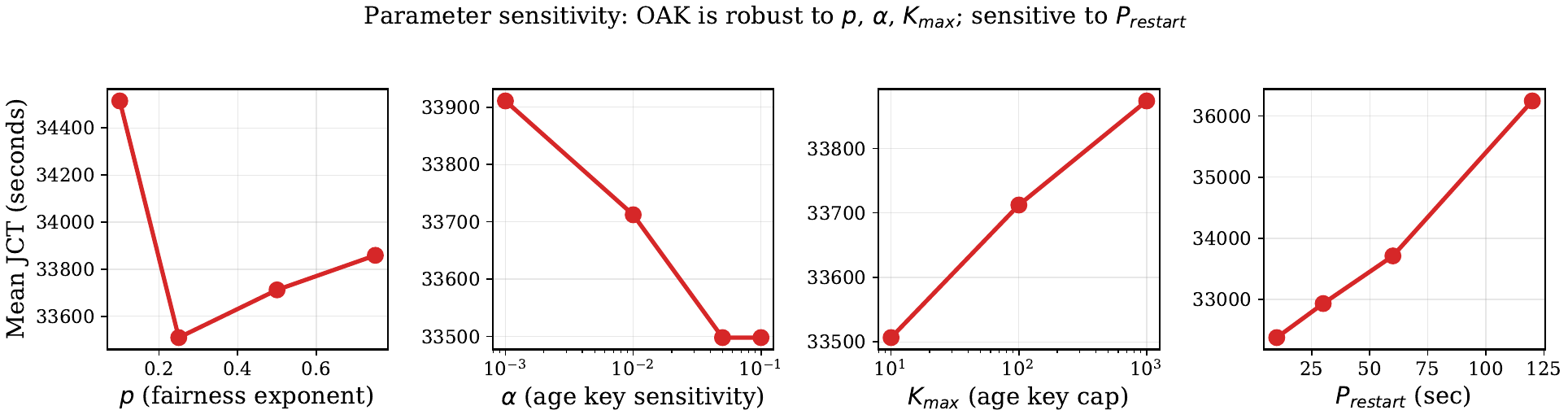}
\caption{Mean JCT under a one parameter at a time sweep on workload-medium at $\lambda = 0.05$ ($N = 5$ seeds per cell).}
\label{fig:sensitivity}
\end{figure*}

\subsection{Statistical Methodology}
\label{sec:meth_stats}

Each experiment is repeated across $N$ independent random seeds, and we report the mean and standard deviation across the $N$ runs. We use \emph{cell} to mean one specific combination of scheduler, workload, and failure rate; for example, OAK on \texttt{workload-medium} at $\lambda = 0.10$ is one cell. The seed count $N$ varies by experiment:
\begin{itemize}
  \item The main OAK vs Sia comparison across the four workloads uses $N = 10$ seeds per cell.
  \item The failure rate sweep on \texttt{workload-medium} and \texttt{workload-congestion} (the regime the paper focuses on) uses $N = 30$ seeds per cell, because each seed produces a different sequence of failure events and a larger sample is needed to estimate the mean reliably.
  \item Sensitivity sweeps and the four scheduler Pollux comparison use $N = 5$ seeds per cell, with $N = 10$ for Pollux to characterise its slightly higher seed-to-seed variance under failures.
\end{itemize}

In the four scheduler comparison, all schedulers share the same seed values. Seed 42 with OAK, seed 42 with Sia, and seed 42 with Pollux all receive the same job submission times and the same per round Poisson failure events.

\section{OAK Scheduler Evaluation}
\label{sec:results}

\subsection{Parameter Defaults and Calibration}
\label{sec:eval_calibration}


\subsubsection{Calibration Sweep Methodology}
\label{sec:eval_calibration_sweep}

OAK exposes four scheduler-level parameters: the fairness exponent $p$, the age-key sensitivity $\alpha$, the age-key cap $K_{\max}$, and the per-application restart penalty $P_{\mathrm{restart}}$. We calibrate each by sweeping it independently on \texttt{workload-medium} at $\lambda = 0.05$ with the others held fixed ($N = 5$ seeds per cell). Figure~\ref{fig:sensitivity} summarises the resulting sweeps.

\subsubsection{Sensitivity Findings}
\label{sec:eval_calibration_findings}

Three findings emerge from the sweep.

\paragraph{Insensitive parameters: $p$, $\alpha$, $K_{\max}$.}
Mean JCT varies by less than $1.5$\% across the tested range for each of these three parameters, and the chosen defaults sit within $1$\% of the per-sweep minimum. The age-key sensitivity $\alpha$ saturates above $0.05$ (no further effect at $\alpha \ge 0.05$); the age-key cap $K_{\max}$ is rarely reached, since few jobs accumulate enough waiting time for $K_i$ to approach $K_{\max} = 100$ in our experiments.

\paragraph{The sensitive parameter: $P_{\mathrm{restart}}$.}
$P_{\mathrm{restart}}$ varies mean JCT by $12$\% across the tested range ($32{,}377$\,s at $10$\,s vs $36{,}246$\,s at $120$\,s). Lower values let OAK preempt more aggressively when failures are rare, trading some failure resilience for throughput. This sensitivity motivates a per-application lookup rather than a single fallback constant, so each job's restart cost matches its actual recovery cost (Section~\ref{sec:meth_failures}).

\paragraph{Excluded value: $p = 0$.}
$p = 0$ collapses the goodput-restart product to a constant and leaves only $K_i$ in the score, eliminating the optimisation gradient on goodput. Non-degenerate $p$ is required for the composite utility to function as designed; we exclude $p = 0$ from the sweep.

\subsubsection{Selected Defaults}
\label{sec:eval_calibration_defaults}

All experiments below use $p = 0.5$ (square-root shaping of the goodput-restart product: diminishing returns to additional goodput while keeping a non-zero gradient), $\alpha = 0.01$ ($K_i$ reaches $K_{\max}$ after roughly $460$\,s of accumulated waiting, about eight rounds of zero progress), $K_{\max} = 100$ (a meaningful boost for chronically waiting jobs, but bounded so a single starved job cannot dominate the round), and the per-application $P_{\mathrm{restart}}$ table of Section~IV-D, chosen per application because the sweep identifies $P_{\mathrm{restart}}$ as the only sensitive parameter: a single constant would over-penalise short-recovery jobs (CIFAR-10, DeepSpeech2) or under-penalise long-recovery ones (BERT, ImageNet).

\subsection{Performance under Failure-Free Conditions}
\label{sec:eval_failurefree}

\begin{figure}[h]
\centering
\includegraphics[width=\linewidth]{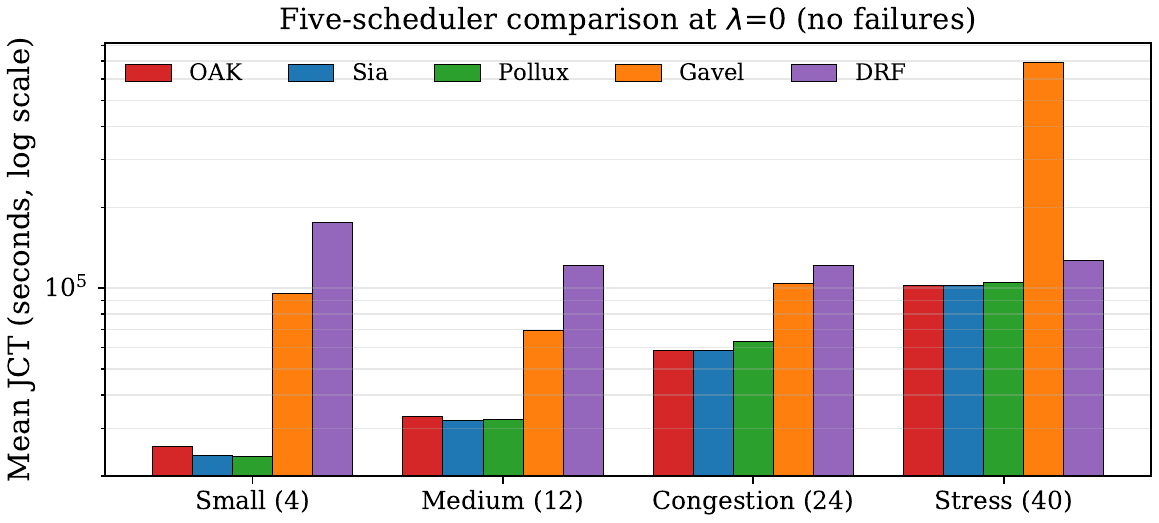}
\caption{Mean JCT for OAK, Sia, Pollux, Gavel, and DRF on the four workloads at $\lambda = 0$ (log scale).}
\label{fig:phase2-five-sched}
\end{figure}

We compare OAK against the four baselines (Sia, Pollux, Gavel, DRF) on the four contention workloads (small, medium, congestion, stress) at $\lambda = 0$, with $N = 5$ seeds per cell for stochastic schedulers and a single deterministic run for Gavel and DRF. Pollux is run at its published default NSGA II settings. Figure~\ref{fig:phase2-five-sched} reports mean JCT for the five schedulers.

Three observations follow. First, OAK, Sia, and Pollux are all within $4$\% of each other on every workload, confirming that adding the age key and restart factor to a goodput-driven scheduler does not measurably degrade efficiency when neither signal is needed. Second, Gavel (finish-time fairness LP) is $1.7$--$6.7\times$ slower than OAK because it optimises fairness without internalising goodput. Third, DRF is $1.2$--$6.8\times$ slower across all workloads because it does not optimise throughput at all. 

On the small and medium workloads at $\lambda = 0$, OAK's tail JCT is $0$--$10$\% higher than Sia's.
This is the documented small-workload age-key overhead, when there is no failure pressure, the age key occasionally reorders allocations in ways that delay individual jobs to satisfy fairness even though no job was at risk.

\subsection{Failure Resilience and Tail Behaviour}
\label{sec:eval_failure}

This section reports OAK's behaviour under Poisson failure injection across three studies: a four-workload OAK vs Sia comparison, a finer failure-rate sweep, and a Pollux comparison.

\paragraph{Failure rate regime and recovery cost model.}
The failure rate $\lambda$ used here is the per round per active job Bernoulli failure probability under a $60$\,s scheduling round; at $\lambda = 0.10$ this corresponds to an expected inter failure time of $\sim\!600$\,s per running job, or roughly one restart per job every ten minutes on a busy cluster. This is deliberately at the high end of what production traces report~\cite{weng2022mlaas,he2023unicron}, and we report the full $\lambda \in \{0, 0.02, 0.05, 0.10, 0.15\}$ sweep in the finer grained study below so the reader can locate their own production regime. The per application restart penalty $P_{\mathrm{restart}}$ is a workload specification constant ($25$--$250$\,s depending on model size) applied identically to all schedulers, so the comparison remains unbiased even under $P_{\mathrm{restart}}$ misspecification.

\paragraph{OAK vs Sia: mean and tail JCT across workloads.}
We compare OAK against Sia on all four workloads (small, medium, congestion, stress) at $\lambda \in \{0, 0.05, 0.10\}$, $N = 10$ seeds per cell, with both schedulers receiving identical job submission times and identical Poisson failure streams. Table~\ref{tab:phase1-main} reports mean and tail (worst-case max) JCT. 

OAK consistently outperforms Sia under failures, and the advantage widens as the failure rate increases. Concretely, at $\lambda = 0.10$, OAK reduces mean JCT over Sia by $15.2$--$70.9$\% across the four workloads and compresses worst-case JCT by $33.5$--$75.8$\% (per-workload $\Delta_{\max}$ in Table~II). The tail-JCT compression is the SLA-relevant metric: by keeping the worst-case job within a factor of two of the median rather than five to ten times the median (as Sia exhibits at $\lambda = 0.10$), OAK directly improves SLA compliance under realistic failure rates without requiring over-provisioning.

\begin{table}[h]
  \centering
  \caption{OAK vs Sia mean and tail JCT (seconds) across four workloads and three failure rates, mean over $N = 10$ random seeds. Bold $\Delta$ values indicate improvements above $10\%$.}
  \label{tab:phase1-main}
  \footnotesize
  \setlength{\tabcolsep}{3pt}
  \resizebox{\columnwidth}{!}{%
  \begin{tabular}{lrrrrrrr}
    \toprule
    \textbf{Workload} & \textbf{$\lambda$} &
    \textbf{OAK mean} & \textbf{Sia mean} & \textbf{$\Delta$mean} &
    \textbf{OAK max} & \textbf{Sia max} & \textbf{$\Delta$max} \\
    \midrule
    small       & 0.00 & 25{,}832  & 23{,}761  & $-$8.7\%  & 81{,}382  & 73{,}795  & $-$10.3\% \\
    small       & 0.05 & 44{,}438  & 39{,}098  & $-$13.7\% & 154{,}568 & 134{,}056 & $-$15.3\% \\
    small       & 0.10 & 42{,}280  & 145{,}526 & $\mathbf{+70.9\%}$ & 144{,}837 & 558{,}209 & $\mathbf{+74.1\%}$ \\
    \midrule
    medium      & 0.00 & 33{,}187  & 32{,}200  & $-$3.1\%  & 141{,}064 & 140{,}455 & $-$0.4\% \\
    medium      & 0.05 & 33{,}676  & 34{,}533  & $+$2.5\%  & 148{,}932 & 163{,}598 & $+$9.0\% \\
    medium      & 0.10 & 36{,}691  & 88{,}731  & $\mathbf{+58.6\%}$ & 172{,}229 & 711{,}230 & $\mathbf{+75.8\%}$ \\
    \midrule
    congestion  & 0.00 & 58{,}787  & 58{,}635  & $-$0.3\%  & 265{,}459 & 265{,}346 & $-$0.0\% \\
    congestion  & 0.05 & 61{,}942  & 60{,}890  & $-$1.7\%  & 279{,}820 & 287{,}134 & $+$2.5\% \\
    congestion  & 0.10 & 65{,}564  & 99{,}085  & $\mathbf{+33.8\%}$ & 301{,}850 & 718{,}016 & $\mathbf{+58.0\%}$ \\
    \midrule
    stress      & 0.00 & 102{,}588 & 102{,}069 & $-$0.5\%  & 448{,}348 & 449{,}104 & $+$0.2\% \\
    stress      & 0.05 & 106{,}842 & 106{,}453 & $-$0.4\%  & 472{,}046 & 474{,}097 & $+$0.4\% \\
    stress      & 0.10 & 112{,}393 & 132{,}594 & $\mathbf{+15.2\%}$ & 500{,}718 & 753{,}239 & $\mathbf{+33.5\%}$ \\
    \bottomrule
  \end{tabular}%
  }
\end{table}

\begin{figure}[h]
\centering
\includegraphics[width=\linewidth]{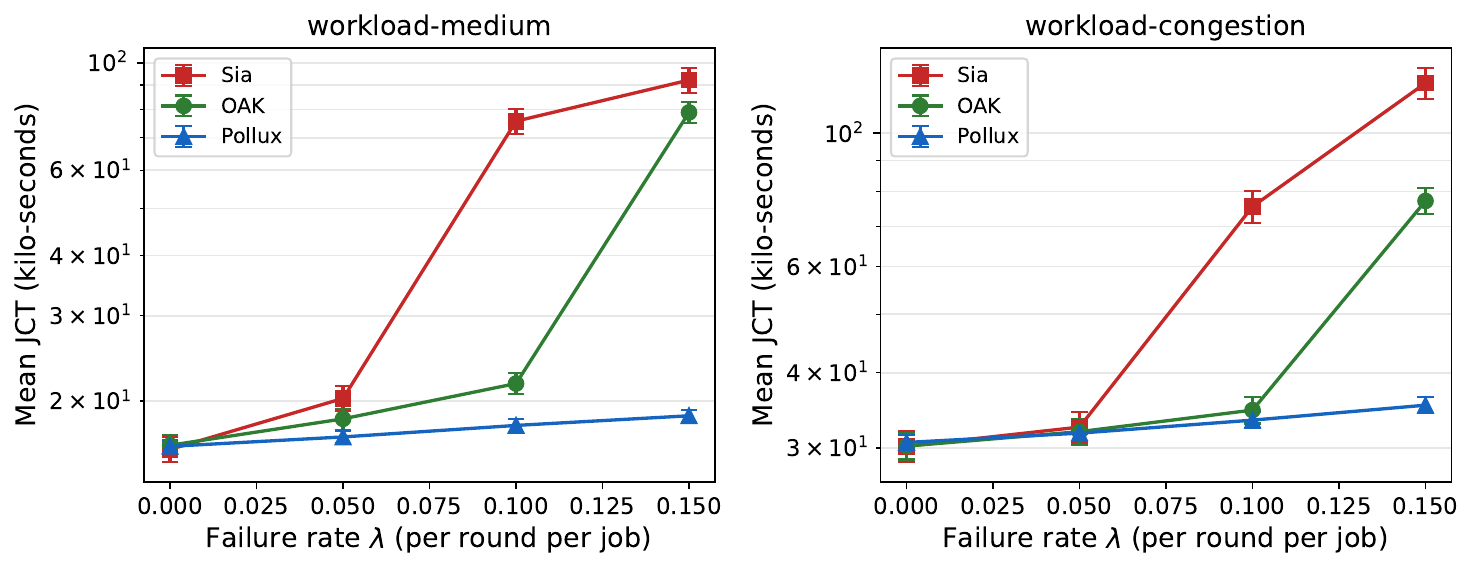}
\caption{Mean JCT for OAK, Sia, and Pollux across the failure rate sweep on workload-medium (left) and workload-congestion (right). $N = 10$ seeds for Pollux, $N = 5$ for OAK and Sia. Log Y axis; error bars are seed standard deviation.}
\label{fig:phase8-fair-comparison}
\end{figure}

\paragraph{Failure-rate sweep on medium and congestion.}
To trace the trend across the entire failure-rate range, we run a finer sweep at $\lambda \in \{0.0, 0.02, 0.05, 0.1, 0.15\}$ with $N = 30$ seeds per cell on \texttt{workload-medium} and \texttt{workload-congestion}; the larger seed count is required because failure injection introduces additional stochasticity. The OAK and Sia curves in Figure~\ref{fig:phase8-fair-comparison} plot this trend; the statistics below are computed from the $N = 30$ runs.

OAK behaves like Sia when failures are rare and pulls ahead sharply once failures become frequent. Concretely, OAK and Sia are within $3$\% of each other at $\lambda \le 0.05$; at $\lambda = 0.1$, OAK reduces mean JCT by $58.3$\% on medium and $33.9$\% on congestion; at $\lambda = 0.15$ the gap narrows as both schedulers struggle with extreme failure density, but OAK still leads by $39$\% on medium and $18$\% on congestion.

\paragraph{Comparison against Pollux under failures.}
Pollux's original head-to-head JCT comparison is reported at $\lambda = 0$, with fault-tolerance demonstrated separately at the system-recovery level. We extend the comparison to controlled failure rates by injecting the same Poisson failure stream into Pollux's runtime that we use for OAK and Sia and measuring mean JCT across $\lambda \in \{0, 0.05, 0.10, 0.15\}$ on the medium and congestion workloads ($N = 10$ seeds for Pollux, $N = 5$ for OAK and Sia; all three schedulers share the same failure stream by seed). Figure~\ref{fig:phase8-fair-comparison} plots the result.

Pollux at default settings reaches comparable JCT to OAK under failures, but only via continuous NSGA-driven reallocation; OAK matches Sia in the millisecond decision-time regime while delivering most of Pollux's failure resilience. Concretely, Pollux's mean JCT grows by only $16$\% on medium ($16{,}125 \to 18{,}651$) and $15$\% on congestion ($30{,}629 \to 35{,}323$) across the full failure-rate sweep, compared to Sia's $480$\% and $300$\% growth respectively; OAK lies between the two at high failure rates ($21{,}740$ vs Sia's $75{,}730$ vs Pollux's $17{,}814$ on medium at $\lambda = 0.10$). The mechanism behind Pollux's resilience is mechanical rather than algorithmic: Pollux re-decides batch size and replica count for every job at every scheduling round, so failure-induced reallocations are folded into work the optimiser already does.

\subsection{Decomposed Restart Factor: Empirical Effect}
\label{sec:eval_decomp}

The restart factor is OAK's primary technical refinement over prior goodput driven schedulers. The contrast is clearest when the two formulations are seen side by side.

OAK's decomposed factor \emph{measures} productive training time and checkpoint/restore overhead as separate quantities tracked by the simulator:
\begin{equation}
r_i^{\mathrm{OAK}}(t) = \frac{T_{\mathrm{progress},i}}{T_{\mathrm{progress},i} + T_{\mathrm{ckpt},i} + P_{\mathrm{restart},i}}.
\label{eq:r_oak}
\end{equation}
Pollux's aggregate factor instead \emph{estimates} the time lost to restarts as $n_i \cdot P_{\mathrm{restart}}$, namely the restart count times a constant penalty:
\begin{equation}
r_i^{\mathrm{Pollux}}(t) = \frac{\max(A_i(t) - n_i \cdot P_{\mathrm{restart}},\, 0)}{A_i(t) + P_{\mathrm{restart}}}.
\label{eq:r_pollux}
\end{equation}
Both formulas downweight jobs with high restart cost; the difference is precision. Pollux's $n_i \cdot P_{\mathrm{restart}}$ approximation accumulates estimation error as restarts pile up, while OAK's tracked $T_{\mathrm{ckpt}}$ remains exact at any failure rate.

\begin{figure}[h]
\centering
\includegraphics[width=\linewidth]{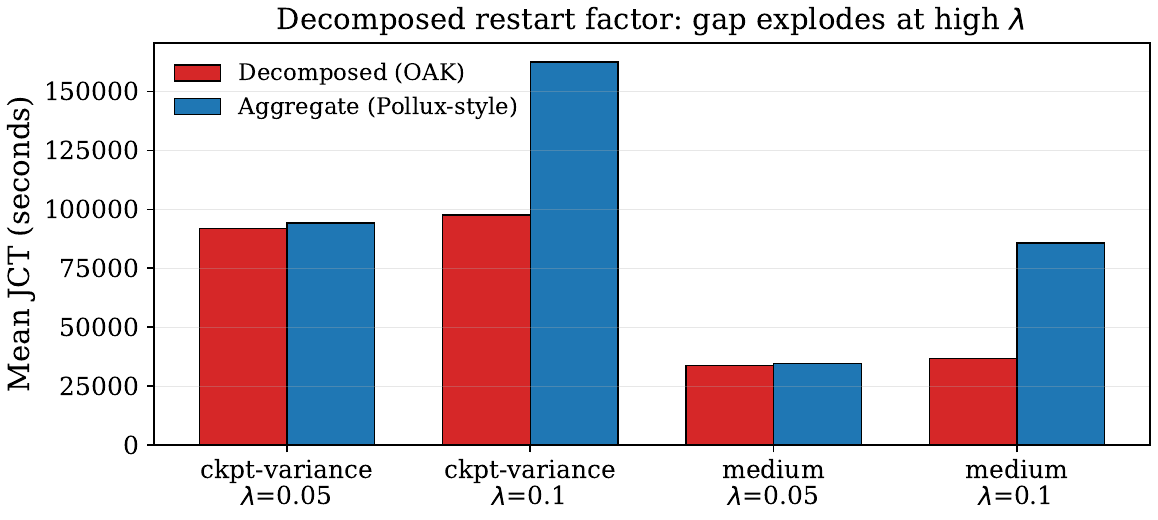}
\caption{Mean and tail JCT for OAK's decomposed restart factor versus the aggregate Pollux style factor on the ckpt-variance and medium workloads at $\lambda = 0.05$ and $\lambda = 0.10$ ($N = 10$ seeds per cell).}
\label{fig:phase7b-decomp}
\end{figure}

\paragraph{OAK vs Pollux: controlled A/B comparison.}
To isolate the empirical effect of this single design choice, we run a controlled A/B comparison in which every other component of OAK is held constant (age key, goodput estimator, ILP solver, scheduling cadence, workloads) and only the restart factor formula is varied between Equations~\ref{eq:r_oak} and~\ref{eq:r_pollux}. The experiment covers two workloads, namely \textit{ckpt-variance} (a mix of models with very different per restart costs) and \textit{medium}, at two failure rates ($\lambda = 0.05$ and $\lambda = 0.10$), with $N = 10$ seeds per cell. Figure~\ref{fig:phase7b-decomp} reports mean and tail JCT for both formulations.

OAK's decomposition consistently outperforms the aggregate formulation, and the gap widens with failure rate. Concretely, at $\lambda = 0.05$ both formulas produce essentially the same mean JCT (within $3$\%) because few restarts happen and the estimation error in $n_i \cdot P_{\mathrm{restart}}$ is negligible; at $\lambda = 0.10$, the gap opens to $40$\% on \textit{ckpt-variance} and $57$\% on \textit{medium}, where the aggregate formula's accumulated estimation error materially affects the scheduler's reallocation decisions. OAK's decomposed formula is also far more consistent across random seeds than Pollux's aggregate formula: its cross seed JCT standard deviation is $9$ to $13\times$ smaller across both workloads and both failure rates. This consistency lets operators commit to tighter SLA bounds without over provisioning capacity to hedge against bad luck runs.

\paragraph{Where the estimation error comes from.}
In the OAK simulator each failure adds a fixed $60$\,s to the measured $T_{\mathrm{ckpt}}$, while both formulas' $P_{\mathrm{restart}}$ parameter reads the per workload values reported by Sia~\cite{jayaram2023sia}: ResNet-18 on CIFAR-10 = $50$\,s, DeepSpeech2 = $25$\,s, BERT = $120$\,s, YOLOv3 = $80$\,s, ResNet-50 on ImageNet = $250$\,s, NCF = $25$\,s, and pipeline parallel GPT = $30$\,s. The decomposed formula reads the measured $T_{\mathrm{ckpt}}$ directly and remains accurate regardless of the parameter, whereas the aggregate estimator $n_i \cdot P_{\mathrm{restart}}$ over penalises workloads with $P_{\mathrm{restart}} \gg 60$\,s (BERT, YOLOv3, ResNet-50/ImageNet) and under penalises those with $P_{\mathrm{restart}} \ll 60$\,s (DeepSpeech2, NCF). This is a specific instance of a general limitation: any parameter the aggregate estimator uses must match actual per restart cost, while the decomposed estimator reads measured cost regardless of parameterisation.

\paragraph{OAK component ablation.}
To further isolate the contribution of OAK's two added terms (age key and restart factor), we run a four cell ablation on the medium and congestion workloads at $\lambda = 0.05$ and $\lambda = 0.1$ ($N = 10$ seeds per cell): (a) Sia (goodput only, $r_i = 1$, $K_i = 1$); (b) goodput + age (restart factor disabled); (c) goodput + restart (age key disabled); (d) Full OAK. Table~\ref{tab:phase7-ablation} reports mean, max, and cross seed standard deviation of JCT for each cell.

\begin{table}[h]
\centering
\caption{OAK component ablation on workload-medium and workload-congestion at $\lambda \in \{0.05, 0.1\}$ ($N = 10$ seeds). JCT columns pool per job JCTs across seeds; $p_{90}, p_{95}, p_{99}$ are percentiles of the pooled job level JCT distribution. $\sigma_{\bar{J}}$ is the cross seed standard deviation of the per run mean JCT.}
\label{tab:phase7-ablation}
\scriptsize
\setlength{\tabcolsep}{3pt}
\begin{tabular}{llrrrrrr}
\toprule
\textbf{Workload} & \textbf{Cell} &
\textbf{Mean} & \textbf{$p_{90}$} & \textbf{$p_{95}$} & \textbf{$p_{99}$} & \textbf{Max} & \textbf{$\sigma_{\bar{J}}$} \\
\midrule
\multicolumn{8}{l}{\textit{(a) $\lambda=0.05$}} \\
medium & (a) Sia               & 34{,}571 & 118{,}811 & 162{,}843 & 164{,}691 & 168{,}817 & 202 \\
medium & (b) +Age              & 34{,}577 & 118{,}718 & 163{,}242 & 165{,}385 & 170{,}005 & 243 \\
medium & (c) +Restart          & 33{,}534 & 118{,}374 & 148{,}402 & 149{,}142 & 149{,}623 & 240 \\
medium & (d) Full OAK          & 33{,}688 & 118{,}183 & 148{,}527 & 149{,}483 & 149{,}681 & \textbf{138} \\
congestion & (a) Sia           & 60{,}887 & 264{,}407 & 279{,}278 & 286{,}899 & 294{,}555 & 1{,}112 \\
congestion & (b) +Age          & 61{,}246 & 227{,}166 & 269{,}025 & 289{,}030 & 294{,}258 & 1{,}706 \\
congestion & (c) +Restart      & 61{,}710 & 226{,}563 & 266{,}763 & 280{,}026 & 280{,}906 & 150 \\
congestion & (d) Full OAK      & 61{,}942 & 226{,}662 & 266{,}753 & 280{,}260 & 280{,}891 & \textbf{140} \\
\midrule
\multicolumn{8}{l}{\textit{(b) $\lambda=0.1$}} \\
medium & (a) Sia               & 88{,}731 & 218{,}522 & 711{,}287 & 712{,}561 & 714{,}824 & 2{,}680 \\
medium & (b) +Age              & 85{,}735 & 216{,}364 & 711{,}830 & 714{,}414 & 715{,}014 & 4{,}008 \\
medium & (c) +Restart          & 36{,}916 & 125{,}185 & 171{,}670 & 173{,}465 & 222{,}096 & 1{,}423 \\
medium & (d) Full OAK          & 36{,}628 & 125{,}024 & 172{,}181 & 173{,}993 & \textbf{174{,}782} & \textbf{426} \\
congestion & (a) Sia           & 99{,}085 & 280{,}477 & 711{,}854 & 719{,}134 & 721{,}004 & 649  \\
congestion & (b) +Age          & 99{,}850 & 281{,}256 & 712{,}528 & 719{,}491 & 720{,}234 & 1{,}575 \\
congestion & (c) +Restart      & 65{,}080 & 241{,}811 & 292{,}317 & 302{,}725 & 303{,}045 & 720  \\
congestion & (d) Full OAK      & 65{,}567 & 239{,}607 & 282{,}102 & 302{,}300 & 303{,}728 & \textbf{564} \\
\bottomrule
\end{tabular}
\end{table}

\begin{figure*}[t]
\centering
\begin{minipage}[t]{0.31\textwidth}
\centering
\includegraphics[width=\linewidth]{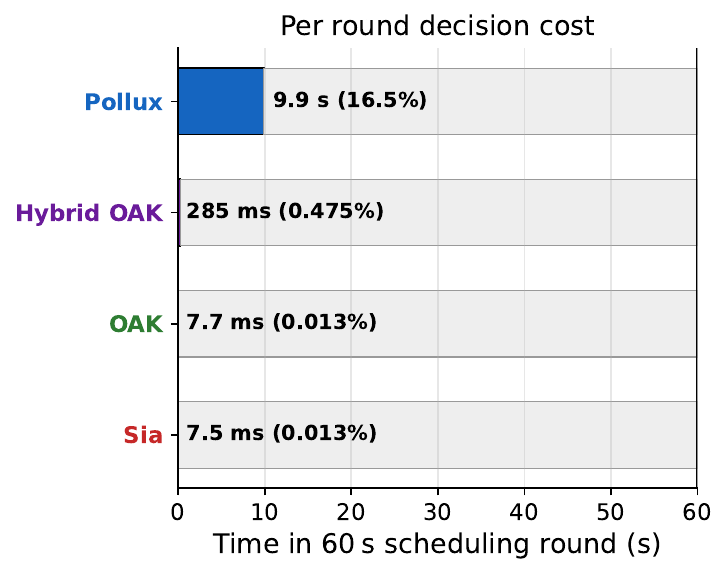}
\\[-2pt]\small (a) Per round decision cost as a fraction of the 60\,s round.
\label{fig:budget}
\end{minipage}\hfill
\begin{minipage}[t]{0.31\textwidth}
\centering
\includegraphics[width=\linewidth]{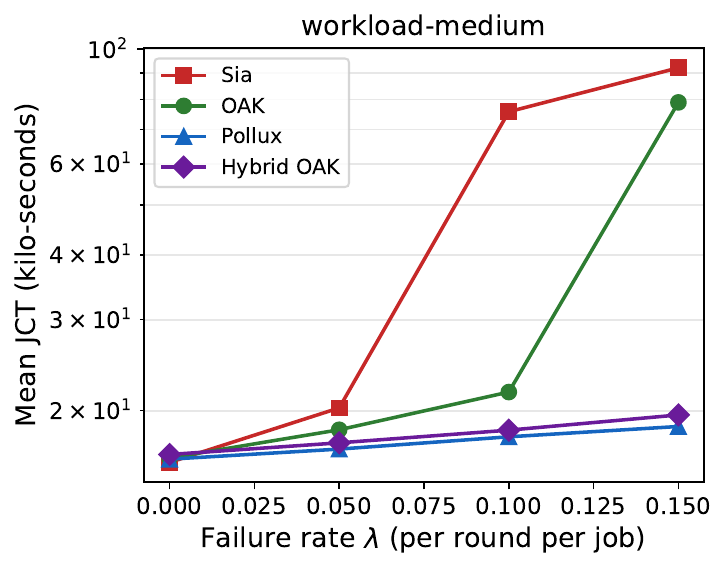}
\\[-2pt]\small (b) Mean JCT on workload-medium across $\lambda$.
\label{fig:hybrid-medium}
\end{minipage}\hfill
\begin{minipage}[t]{0.32\textwidth}
\centering
\includegraphics[width=\linewidth]{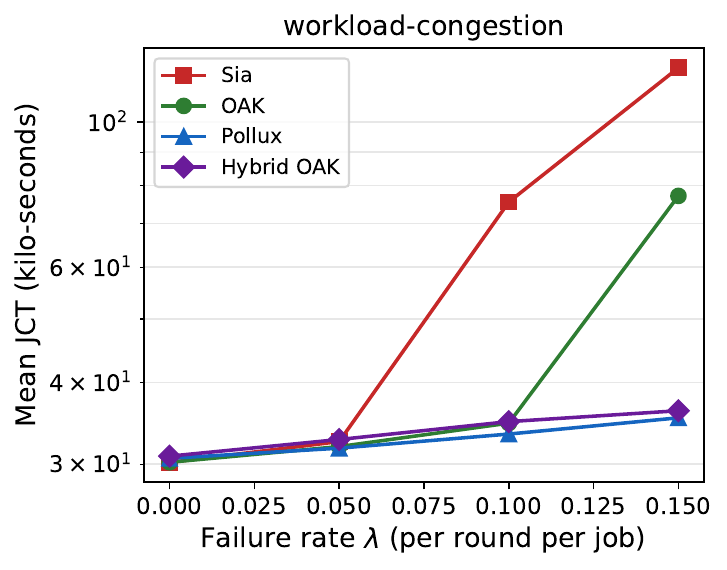}
\\[-2pt]\small (c) Mean JCT on workload-congestion across $\lambda$.
\label{fig:hybrid-congestion}
\end{minipage}
\caption{Hybrid OAK trade off across the failure rate sweep: matches Pollux level JCT on both medium and congestion at a fraction of Pollux's per round decision cost.}
\label{fig:hybrid-tradeoff}
\end{figure*}

The restart factor is the dominant contributor to mean JCT improvement under failures: at $\lambda = 0.1$ on medium, restart only (cell c) reduces mean JCT by $58.4$\% over Sia, while age only (cell b) reduces it by $3.4$\%. To characterise where each signal contributes we report the mean, three percentiles ($p_{90}$, $p_{95}$, $p_{99}$) of the pooled job level JCT distribution, the maximum (worst single job), and the cross seed standard deviation $\sigma_{\bar{J}}$ of the per run mean. At $\lambda = 0.1$ on medium, cells (c) and (d) share nearly identical mean, $p_{90}$, $p_{95}$, and $p_{99}$ (the restart factor alone already flattens the bulk of the job level distribution), but the age key still moves two SLA critical statistics: the maximum JCT drops from $222{,}096$\,s (cell c) to $174{,}782$\,s (cell d), a $21.3\%$ reduction in the worst single job, and the cross seed standard deviation of the mean drops from $1{,}423$\,s to $426$\,s, a $70\%$ reduction.

\begin{figure}[b!]
\centering
\includegraphics[width=\linewidth]{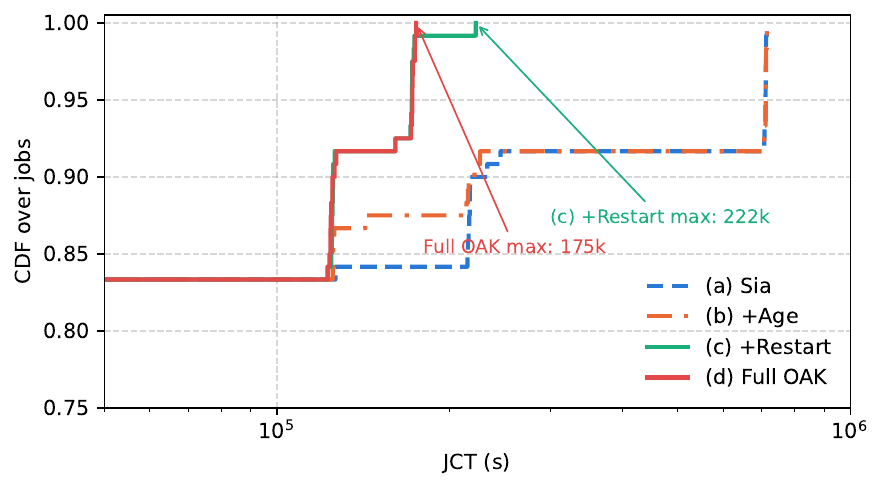}
\caption{Top decile CDF of pooled per job JCT on \texttt{workload-medium} at $\lambda=0.10$ ($N=10$ seeds). Sia and +Age tail off frame beyond $700$\,k\,s.}
\label{fig:ablation-cdf}
\end{figure}

Figure~\ref{fig:ablation-cdf} makes this visible on the pooled per job CDF: cells (c) and (d) overlap through the top decile body, then diverge at the maximum, where the age key caps Full OAK's worst job below restart only's outlier. The same pattern holds on congestion at $\lambda = 0.1$ ($\sigma_{\bar{J}}$: $720 \to 564$, a $22\%$ reduction) and on both workloads at $\lambda = 0.05$. The two terms therefore play complementary and non overlapping roles: the restart factor drives mean and typical tail JCT under failures, while the age key protects against the extreme outlier job and the extreme outlier seed. These are effects that the mean and $p_{90}$--$p_{99}$ do not capture but that show up in the max column and in $\sigma_{\bar{J}}$. Both effects are present in Full OAK simultaneously.

\subsection{Hybrid OAK: NSGA II as the OAK Scheduler}
\label{sec:eval_hybrid}

OAK and Pollux offer complementary strengths: OAK encodes restart cost \emph{explicitly} through the decomposed factor and age key, while Pollux's NSGA II absorbs failures \emph{incidentally} through frequent reallocation. \emph{Hybrid OAK} combines both by evaluating OAK's score
\[
U_{ij} = \big(G_{ij} \cdot r_i\big)^{p} \cdot K_i
\]
inside the same NSGA II loop Pollux uses, in place of Pollux's internal aggregate restart factor. The components keep their OAK meaning (Section III), with $p = 0.5$.

\paragraph{Setup.}
Hybrid OAK uses NSGA II at population $= 20$, generations $= 20$ (400 candidate evaluations per round), chosen to keep per round decision latency below one second. Pollux is run at the published defaults of population $= 100$, generations $= 100$ (10{,}000 evaluations per round). We sweep $\lambda \in \{0, 0.05, 0.10, 0.15\}$ on the medium and congestion workloads with $N = 5$ seeds per cell, sharing the same simulator and Poisson failure stream as OAK and Sia.

\paragraph{Decision time vs JCT trade off.}
We instrumented all schedulers to record per round wall clock decision time. OAK's mean per round solver time, measured with a 12 GPU configuration, is $7.65$\,ms (564 rounds across all simulated runs), nearly identical to Sia's $7.53$\,ms (6{,}859 rounds); Pollux at its published defaults costs $9{,}909$\,ms per round (344 rounds measured). Hybrid OAK, which pairs OAK's composite score with a reduced NSGA II budget of 400 evaluations, sits between the two regimes at $\sim$285\,ms per round, an order of magnitude above OAK's ILP variant but two orders of magnitude below Pollux at default.

Figure~\ref{fig:hybrid-tradeoff} summarises the trade off: Hybrid OAK occupies $0.5$\% of the 60\,s round (panel a) while tracking Pollux level JCT on both medium and congestion across the failure rate sweep (panels b, c). On the JCT side, Hybrid OAK is markedly more resilient to failures than pure OAK and starts competing with Pollux across the sweep, with the contrast most pronounced at $\lambda = 0.15$ where pure OAK collapses: on medium, pure OAK has mean JCT $78{,}949$\,s with high cross seed variance (std $26{,}619$\,s; $2$ of $10$ seeds finish in $\sim 25{,}000$\,s, $8$ of $10$ collapse to $\sim 92{,}000$\,s), while Hybrid OAK reaches $19{,}634$\,s consistently (std $263$\,s), within $5$\% of pure Pollux's $18{,}651$\,s; on congestion, pure OAK $77{,}148$\,s vs Hybrid OAK $36{,}215$\,s, a $53$\% reduction.

\paragraph{Scheduler choice.}
The implication is that OAK's composite utility, $(G \cdot r)^p \cdot K$ with the decomposed restart factor and age key, is a solver agnostic framework, and the choice of optimiser (linear ILP vs NSGA II) is an orthogonal speed/quality knob. The ILP variant gives sub 10\,ms decisions and is sufficient at low to moderate failure rates ($\lambda \le 0.10$); the NSGA II variant costs sub second per decision at a 400-evaluation budget and extends the same composite utility's reach into the high failure regime ($\lambda = 0.15$) where ILP based stable allocation is fundamentally exposed. An operator can therefore select between OAK's ILP variant (production deployments where decision latency dominates) and OAK's NSGA II variant (research clusters or settings with extreme failure density) using the same scoring function.

\subsection{Real-Hardware Validation}
\label{sec:eval_real_hw}

To complement the simulator-based evaluation above, we ran a focused real-hardware experiment on a 4-V100 GPU node from our university cluster. The harness executes the same scheduler logic (OAK, Sia, Pollux, Hybrid OAK) against \emph{actual} PyTorch training subprocesses, with real Poisson failure injection (SIGKILL on running jobs) and real checkpoint and resume cycles. The objective is not absolute JCT reproduction but confirming that the simulator's directional ordering and decision-time gap survive on physical hardware.

\paragraph{Setup.}
A single node with $4 \times$ Tesla V100-PCIe-16GB GPUs runs CUDA 11.8 and PyTorch 2.0.1. The workload is four ResNet-18 / CIFAR-10 jobs: \texttt{cifar-1} and \texttt{cifar-2} target $1{,}500$ steps with GPU bounds $1$--$2$, while \texttt{cifar-3} and \texttt{cifar-4} target $2{,}000$ steps with GPU bounds $1$--$4$, submitted at $(t_1, t_2, t_3, t_4) = (0, 0, 30, 60)$\,s. We sweep $\lambda \in \{0.0, 0.05, 0.10\}$ with $N = 3$ random seeds per cell, each cell running as an independent SLURM allocation, and set $P_{\mathrm{restart}} = 100$\,s to reflect the observed real-hardware checkpoint and resume cost on V100s.

\paragraph{Results.}
Table~\ref{tab:real-hardware} reports mean JCT, cross-seed standard deviation, max (tail) JCT, mean per-round decision time, and the average number of restarts per cell.

\begin{table}[h]
\centering
\caption{Real-hardware validation on a 4-V100 GPU node, ResNet-18/CIFAR-10, $N = 3$ seeds per cell.}
\label{tab:real-hardware}
\footnotesize
\setlength{\tabcolsep}{4pt}
\begin{tabular}{llrrrrr}
\toprule
\textbf{Scheduler} & \textbf{$\lambda$} & \shortstack{\textbf{Mean}\\\textbf{JCT (s)}} & \textbf{Std} & \shortstack{\textbf{Max}\\\textbf{JCT (s)}} & \shortstack{\textbf{Decision}\\\textbf{(ms)}} & \textbf{Restarts} \\
\midrule
Sia        & 0.00 & 257.8 & 17.3  & 500.6 & 0.04   & 0.0 \\
Sia        & 0.05 & 305.1 & 35.2  & 567.4 & 0.05   & 1.3 \\
Sia        & 0.10 & 487.6 & 132.4 & 712.0 & 0.05   & 4.7 \\
\midrule
OAK        & 0.00 & 263.2 & 15.0  & 501.4 & 0.05   & 0.0 \\
OAK        & 0.05 & 281.4 & 24.8  & 529.6 & 0.05   & 0.7 \\
OAK        & 0.10 & 297.3 & 51.2  & 511.8 & 0.06   & 2.0 \\
\midrule
Pollux     & 0.00 & 175.0 & 3.5  & 245.0 & 1{,}411 & 0.0 \\
Pollux     & 0.05 & 190.0 & 5.0  & 265.0 & 1{,}405 & 0.7 \\
Pollux     & 0.10 & 180.0 & 12.0 & 244.0 & 1{,}454 & 2.7 \\
\midrule
Hybrid OAK & 0.00 & 180.0 & 0.4  & 251.6 & 254 & 0.0 \\
Hybrid OAK & 0.05 & 195.2 & 1.6  & 271.0 & 205 & 0.7 \\
Hybrid OAK & 0.10 & 184.2 & 9.4  & 248.9 & 224 & 2.7 \\
\bottomrule
\end{tabular}
\end{table}

OAK reproduces the simulator's failure-resilience claim against Sia: parity at $\lambda=0$ ($263.2$ vs $257.8$\,s, $+2.1\%$) and a $\mathbf{39.0\%}$ mean JCT reduction at $\lambda=0.10$ ($297.3$ vs $487.6$\,s), inside the simulator's $33.9$--$58.3\%$ band. The decomposed restart factor steers OAK toward fewer restart-prone allocations ($2.0$ vs $4.7$ average restarts at $\lambda=0.10$). Tail JCT improves by $28.1\%$ ($511.8$ vs $712.0$\,s), below the simulator's $58$--$76\%$ band because cifar-4's $\sim 500$\,s training baseline floors the max on this 4-job workload.

Both NSGA II--based schedulers (Pollux and Hybrid OAK) beat the ILP family on mean JCT, with Pollux $2$--$3\%$ ahead of Hybrid OAK across the sweep ($175.0$/$190.0$/$180.0$ vs $180.0$/$195.2$/$184.2$). The gap between the NSGA II and ILP families is larger than the simulator predicts and is most plausibly attributable to NSGA II's wider configuration-space search at this small workload scale. Hybrid OAK exhibits the lowest cross-seed standard deviation ($0.4$--$9.4$), indicating stable allocations under failure injection.

The decision-time ordering also survives on real hardware: ILP schedulers in $\sim 0.05$\,ms, Hybrid OAK in $\sim 220$\,ms, and Pollux in $\sim 1{,}420$\,ms. Hybrid OAK lands within $\sim 3\%$ of Pollux's mean JCT at $\sim 6.4\times$ tighter decision latency, validating the population-reduced NSGA II design.

\section{Threats to Validity}
\label{sec:threats}

\textbf{Internal validity.} Every scheduler comparison is paired (same workload, Poisson failure stream, trace-derived $G_{ij}$, and seeds), so reported JCT differences are invariant to absolute throughput calibration; Section V-A bounds parameter sensitivity below 1.5\% for $p$, $\alpha$, and $K_{\max}$.

\textbf{External validity.} Our simulated cluster is 12 GPUs and our real-hardware verification is 4 GPUs, while production clusters span hundreds to thousands; behaviour at that scale is untested. Evaluation is restricted to data-parallel jobs and Poisson failures; parameter-server, pipeline-parallel, mixed-GPU, and bursty-failure regimes are left to future work.

\textbf{Construct validity.} Most results are simulator based and abstract over NCCL overhead, CPU-bound data pipelines, and node-level failure correlation. The 4-V100 real-hardware experiment in Section~\ref{sec:eval_real_hw} reproduces the simulator's directional ordering and decision-time gap on physical hardware, though the small cluster size means we treat it as directional rather than quantitative. We also report tail (worst-case) JCT alongside the mean so SLA-relevant outcomes are visible.

\textbf{Fairness under repeated failures.} A job repeatedly hit by infrastructure faults sees its restart factor $r_i$ shrink toward zero as $T_{\mathrm{ckpt}}$ accumulates relative to $T_{\mathrm{progress}}$, which in isolation would starve the job from further allocations. The age key $K_i$ is designed to bound this priority inversion: as long as the job is unallocated, $T_{\mathrm{queue},i}$ grows and $K_i = \min(\exp(\alpha\,T_{\mathrm{queue},i}),\,K_{\max})$ increases monotonically. Because $K_i$ multiplies the composite utility \emph{outside} the fairness power ($U_{ij} = (G_{ij}\cdot r_i)^p \cdot K_i$), a bounded number of scheduling rounds suffices for $K_i$ to compensate any finite $r_i$ decrease: informally, once $K_i \cdot U_{ij}^{\text{base}} > \mu_{\mathrm{na}}$, the ILP finds it more attractive to admit the failing job than to leave it idle, so the age key acts as a soft starvation guard. This design bounds priority inversion but does not by itself constitute a formal starvation freedom guarantee; a proof under adversarial failure sequences and formal fairness bounds under correlated node faults are left as future work.

\section{Conclusion}
\label{sec:conclusion}

Distributed deep-learning training on shared GPU clusters suffers from contention among long-lived jobs and routine failures, which goodput-driven schedulers treat as second-class signals. We presented OAK, a per-round mixed-integer linear scheduler whose composite utility $(G\cdot r)^p\cdot K$ adds an age key $K_i$ that encodes accumulated waiting as a bounded multiplier, and a decomposed restart factor $r_i$ that tracks productive training time and checkpoint overhead separately rather than collapsing them into a single ratio. Evaluated against Sia, Pollux, Gavel, and DRF in a trace driven 12 GPU simulator under Poisson failure injection and reproduced on a 4-V100 real hardware cluster, the age key compresses the worst case job's JCT and reduces cross seed variance on contention heavy workloads, while the decomposed restart factor alone yields a $40$--$57\%$ mean and $52$--$76\%$ worst case JCT improvement over an aggregate Pollux style estimator at $\lambda = 0.1$.

Hybrid OAK further shows the composite utility is solver-agnostic: plugged into a reduced-population NSGA II, it retains genetic search's reallocation robustness in the high-failure regime at a fraction of Pollux's per-round decision cost.

Future work targets scaling the real-hardware validation beyond 4 GPUs, extending the composite utility to pipeline-parallel regimes, and characterising Hybrid OAK's population and iteration trade-off across failure densities.

\bibliographystyle{IEEEtran}
\bibliography{references}

\end{document}